# Functional photoacoustic noninvasive Doppler angiography in humans

Yang Zhang[1†], Joshua Olick-Gibson [1†], Karteekeya Sastry[2], Lihong V. Wang[1,2,*]

[1] Caltech Optical Imaging Laboratory, Andrew and Peggy Cherng Department of Medical Engineering, California Institute of Technology, 1200 East California Boulevard, Pasadena, CA 91125, USA.

[2] Caltech Optical Imaging Laboratory, Department of Electrical Engineering, California Institute of Technology, 1200 East California Boulevard, Pasadena, CA 91125, USA.

[†]These authors contributed equally to this work: Yang Zhang, Joshua Olick-Gibson

*Corresponding author. L. V. Wang (LVW@caltech.edu)

## Abstract

Optical imaging of blood flow yields critical functional insights into the circulatory system, but its clinical implementation has typically been limited to shallow depths (~1 millimeter) due to light scattering in biological tissue. Here, we present photoacoustic noninvasive Doppler angiography (PANDA) for deep blood flow imaging. PANDA synergizes the photoacoustic and Doppler effects to generate color Doppler velocity and power Doppler blood flow maps of the vascular lumen. Our results demonstrate PANDA's ability to measure blood flow *in vivo* up to one centimeter in depth, marking approximately an order of magnitude improvement over existing high-resolution pure optical modalities. PANDA enhances photoacoustic flow imaging by increasing depth and enabling cross-sectional blood vessel imaging. We also showcase PANDA's clinical feasibility through three-dimensional imaging of blood flow in healthy subjects and a patient with varicose veins. By integrating the imaging system onto a mobile platform, we have designed PANDA to be a portable modality that is primed for expedient clinical translation. PANDA offers noninvasive, single modality imaging of hemoglobin and blood flow with three-dimensional capability, facilitating comprehensive assessment of deep vascular dynamics in humans.

## Introduction

As a hallmark of cardiovascular, neurological, and oncological function, blood flow quantification plays a crucial role in biomedical imaging for physiologists and clinicians alike. For several decades, Doppler-based methods have been a mainstay in biomedical optics due to their robust implementation in quantifying blood flow. Laser Doppler velocimetry (LDV) and Doppler optical coherence tomography (OCT) are two widely used methods for blood flow characterization. LDV detects frequency shifts by collecting scattered light from particles moving through a laser-induced interference fringe pattern[1], whereas Doppler OCT detects a frequency shift between backscattered light and a reference beam to form high-resolution tomographic images of vasculature hemodynamics[2]. Despite significant developments in these techniques over the years, they are limited to depths in biological tissue less than the optical diffusion limit (~1 mm) due to strong scattering attenuation [3].

In addition to optical imaging modalities, other techniques such as Doppler ultrasound and magnetic resonance imaging (MRI) have also proven effective in measuring blood flow. Pulsed-echo Doppler ultrasound can image deep blood flow at centimeter-level depths in the body, relying on the Doppler effect. One notable implementation is functional ultrasound, which enables blood flow measurement for evaluating brain function[4,5]. However, ultrasound imaging does have its limitations, particularly in lacking the ability to provide molecular contrast, such as hemoglobin content and oxygen saturation, both of which are crucial physiological parameters for understanding hemodynamics. On the other hand, MRI has also been utilized for measuring deep blood flow; nonetheless, it does have some drawbacks, including limited spatial resolution and typically higher costs associated with its usage[6,7].

Photoacoustic tomography (PAT) is uniquely positioned in the field of biomedical imaging. As a hybrid modality that converts light into sound via thermoelastic expansion, PAT enables optical absorption contrast with acoustic resolution[3]. In its original conception, the photoacoustic Doppler (PAD) effect was achieved by positioning an intensity-modulated laser obliquely to an ultrasonic transducer[8]. Subsequent implementations of the PAD effect have been explored at both the optical and acoustic resolution scales. Optical resolution photoacoustic microscopy (OR-PAM) employs bandwidth broadening and bidirectional scanning to obtain flow quantification and direction[9]. More recently, OR-PAM has been employed to quantify blood flow in the superficial mouse brain[10] and mouse ear[11]. However, OR-PAM, like other high-resolution pure optical techniques, has a limited imaging depth due to the inability to focus light beyond one millimeter in biological tissue[12–15]. Conversely, acoustic resolution photoacoustic Doppler flowmetry (AR-PAF) has demonstrated the ability to measure flow in blood-mimicking[16] and whole blood[17] phantoms using a single element transducer and wide-field illumination by computing A-line cross-correlations between photoacoustic waveforms. Nonetheless, AR-PAF studies have been only confined to *in vitro* and *ex vivo* measurements. In photoacoustic computed tomography (PACT), previous papers have employed ultrasonically encoded photoacoustic flowgraphy to track propagating thermal nonuniformities induced by ultrasonic heating, but in all of these studies, blood flow measurements were relegated to *ex vivo* scenarios[18]. Additionally, cuff-induced *in vivo* blood flow measurements have been explored in deep tissue[19], but these methods were only capable of transient measurements, while naturally occurring physiological flow measurements have remained unachievable.

This study introduces photoacoustic noninvasive Doppler angiography (PANDA) as an innovative technique for acquiring the first known photoacoustic Doppler maps of human blood flow by breaking through the optical diffusion limit. Notably, PANDA exhibits the capability to characterize blood flow in vessels at depths reaching one centimeter, signifying an advancement over existing high-resolution optical methods[14,15] by nearly an order of magnitude, and unlocking the potential to utilize PANDA in medically significant scenarios. To validate the accuracy of PANDA, color Doppler and power Doppler measurements in *ex vivo* blood were compared against preset syringe flow speeds. The applicability of PANDA was further demonstrated *in vivo*, capturing blood flow images of vessels up to one centimeter deep in human subjects. By linearly scanning our ultrasonic probe and optical fiber bundle, we obtained three-dimensional PANDA images to provide volumetric hemodynamic characterization of the human vasculature. Additionally, we developed a matrix array-based implementation of PANDA to acquire fast three-dimensional *in vivo* blood flow maps from the carpal tunnel and dorsal hand regions. To monitor functional changes, we implemented PANDA to image the dynamics of *in vivo* blood flow responses induced by the inflation and subsequent release of a blood pressure cuff. Furthermore, to demonstrate the clinical viability of PANDA, we acquired a three-dimensional map of blood flow from the foot of a patient presenting with varicose veins. To facilitate PANDA's transition into clinical practice, we integrated the imaging components onto a mobile platform. As a portable and noninvasive imaging modality, PANDA holds promising potential for future clinical utilization, particularly in the realm of bedside hemodynamic evaluation for patient care.

**Results**

The framework for photoacoustic noninvasive Doppler angiography (PANDA) is explained in Fig. 1. As shown by the system schematic in Fig. 1a, a laser delivers light through an optical fiber bundle coupled to an ultrasonic probe. The light is absorbed by the hemoglobin molecules in the blood vessels, after which acoustic waves are generated through thermoelastic expansion. The acoustic waves are then detected by the ultrasonic probe and streamed from a data acquisition (DAQ) device to a host computer for image reconstruction using the universal back-projection algorithm[20]. To make the system portable, we integrated the aforementioned components onto a mobile platform (see Supplementary Fig. 1). Because the ultrasonic probe can only form single-shot, two-dimensional images, we also synchronized the laser and DAQ device with a motorized translation stage and linearly scanned the probe along its elevational direction to form three-dimensional (3D) images. In order to detect the blood movement, it is necessary for the blood lumen photoacoustic signals to not be suppressed relative to the blood boundary signals. We hypothesize that this condition is satisfied by a non-uniform random distribution of photoacoustic absorbers (i.e., the hemoglobin content of the red blood cells (RBCs)), as illustrated in Fig. 1b. Fig. 1c shows a representative photoacoustic computed tomography (PACT) image in which the blood lumen signals are still detectable relative to the strong boundary signals. After acquiring successive, wide-field photoacoustic images, temporal frequency analysis (Fig. 1d, top panel) on representative pixels in the blood and tissue regions of the reconstructed image in Fig. 1c show a strong low-frequency component in the static tissue region and a high-frequency component in the fast-moving blood region. The blood signals can be extracted by applying a bandpass filter to eliminate contributions from the low-frequency tissue and high-frequency noise components. Next, the Doppler frequency shift ($f_D$) from the blood can be estimated (Fig. 1d, bottom panel) and used to calculate the blood flow velocity ($v$) according to $v = \frac{f_D}{f_0} \frac{c}{\cos\theta}$, where $f_0$ is the probe's center-frequency, $c$ is the speed of sound in the medium, and $\theta$ is the angle between the axis of the probe and the longitudinal axis of the blood vessel. Performing the above analysis on each pixel of the image generates a color Doppler map, as shown in Fig. 1e, where $v_{axial}$ represents the axial component (relative to the probe surface, as shown by the $z$ axis in Fig. 1a) of the blood flow velocity ($v_{axial} = v \cos\theta$). Moreover, we can generate a power Doppler map (see Fig. 1f) by calculating the pixelwise mean blood signal intensities, which are proportional to the blood volume. Further signal processing details can be found in the Methods section.

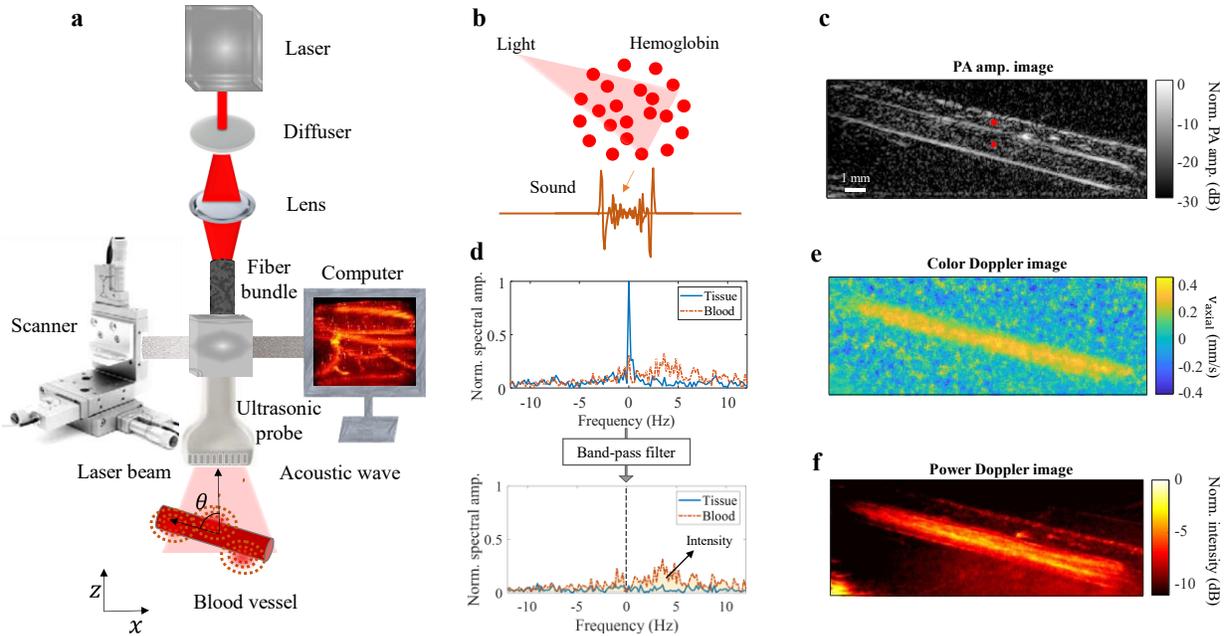

**Fig. 1 | Principle of photoacoustic noninvasive Doppler angiography (PANDA). a,** System setup for PANDA. Light passes through a fiber bundle and is delivered to a blood vessel. An ultrasonic probe detects the emitted photoacoustic waves, and the raw data is sent to a host computer for reconstruction. The probe is mounted to a scanner in order to obtain 3D images. **b,** Lumen signal visibility. A nonuniform random distribution of photoacoustic absorbers (i.e., hemoglobin) gives rise to increased lumen signal amplitudes relative to the strong blood boundary signals. **c,** PACT structural image of a blood vessel, which is located in the forearm region. The upper and lower red markings indicate representative pixels in the tissue and blood vessel lumen regions, respectively. **d,** Frequency analysis of PACT image stack before (top) and after (bottom) applying a bandpass filter. The bandpass filter removes the low-frequency tissue component as well as the high-frequency background noise. **e,** Representative color Doppler velocity image. **f,** Representative power Doppler image.

To assess the validity of PANDA, we constructed a bifurcating blood flow phantom in which two inlet channels converged into an outlet channel (see Fig. 2a). The structural photoacoustic images for the outlet channel are shown for longitudinal and cross-sectional views in Fig. 2b and 2c, respectively. Longitudinal and cross-sectional images of blood flow in the outlet channel were acquired at five syringe flow speeds ranging from 0.5-4.0 mm/s. As shown by the longitudinal color Doppler maps in Fig. 2d and the corresponding plot in Fig. 2g, the axial velocity measurements become proportionally more negative with increasing syringe flow speed; here, the negative sign is consistent with the true blood flow direction in this phantom moving away from the probe (see vector orientation in Fig. 2a). Moreover, the longitudinal and cross-sectional power

Doppler maps in Fig. 2e and 2f, respectively show a similar trend, with both corresponding plots in Fig. 2h and 2i confirming that the power Doppler measurements proportionally increase with syringe flow speed (it also indicates the blood volume). Conversely, the results in Supplementary Fig. 2 show that the photoacoustic amplitude from the structural images does not change with syringe flow speed.

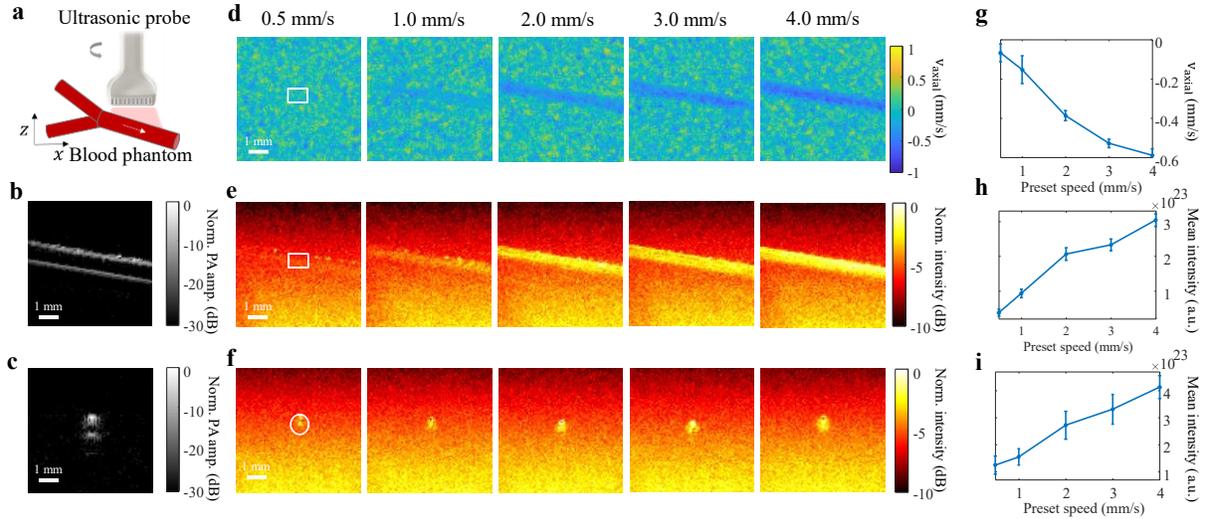

**Fig. 2 | PANDA phantom validation. a,** Schematic of the experimental setup. Blood is perfused through a converging bifurcation to create spatial heterogeneity (to be discussed). Longitudinal and cross-sectional images are acquired on the outlet channel. **b-c,** Structural PACT images in the longitudinal and cross-sectional orientations, respectively. **d-f,** Longitudinal color Doppler, longitudinal power Doppler, and cross-sectional power Doppler images, respectively, measured across five preset syringe flow speeds ranging from 0.5-4.0 mm/s. **g-i,** Mean and standard error values for the white outlined regions of interest in **d-f,** respectively. All three image sets show a proportional relationship between the preset speeds and the measured values.

Six representative *in vivo* PANDA images of vessels at depths beyond the optical diffusion limit are shown in Fig. 3. Fig. 3a-f shows the PACT images, with amplitude values corresponding to hemoglobin-based optical absorption contrast, whereas Fig. 3g-l shows the power Doppler images, which corresponds to the blood volume. Fig. 3a shows three in-plane vessel branches, with the corresponding power Doppler image shown in Fig. 3g. In Fig. 3b, the structural image shows a strong shallow vessel and a weaker deep vessel below it, whereas the power Doppler image in Fig. 3h shows clear delineation of both vessels, with the lumen regions having higher blood flow contrast than the background. Next, a digital artery (shown in Fig. 3c) was imaged while applying a blood pressure cuff to the brachial artery (proximal to the imaging site). The corresponding power Doppler image in Fig. 3i shows that after inflating the cuff, the blood flow in the artery can be clearly visualized. Furthermore, the PANDA image reveals the cross-section of another vessel above the artery, which is not clearly visualized in the structural image. To image vessels at greater depths, we employed a probe with a center frequency of 5 MHz. The structural and power Doppler images for three vessels imaged with this probe are shown in Fig 3d-f and Fig. 3j-l, respectively. The corresponding profiles in Fig. 3p-r reveal that we were able to image blood flow at depths of 4, 5 and 10 millimeters, respectively. As shown in Fig. 3m-r, the PACT structural profiles for all six images display strong blood boundary signals and weaker blood lumen signals, whereas the power Doppler profiles provide better vessel delineation by filling in the lumen regions with blood flow contrast.

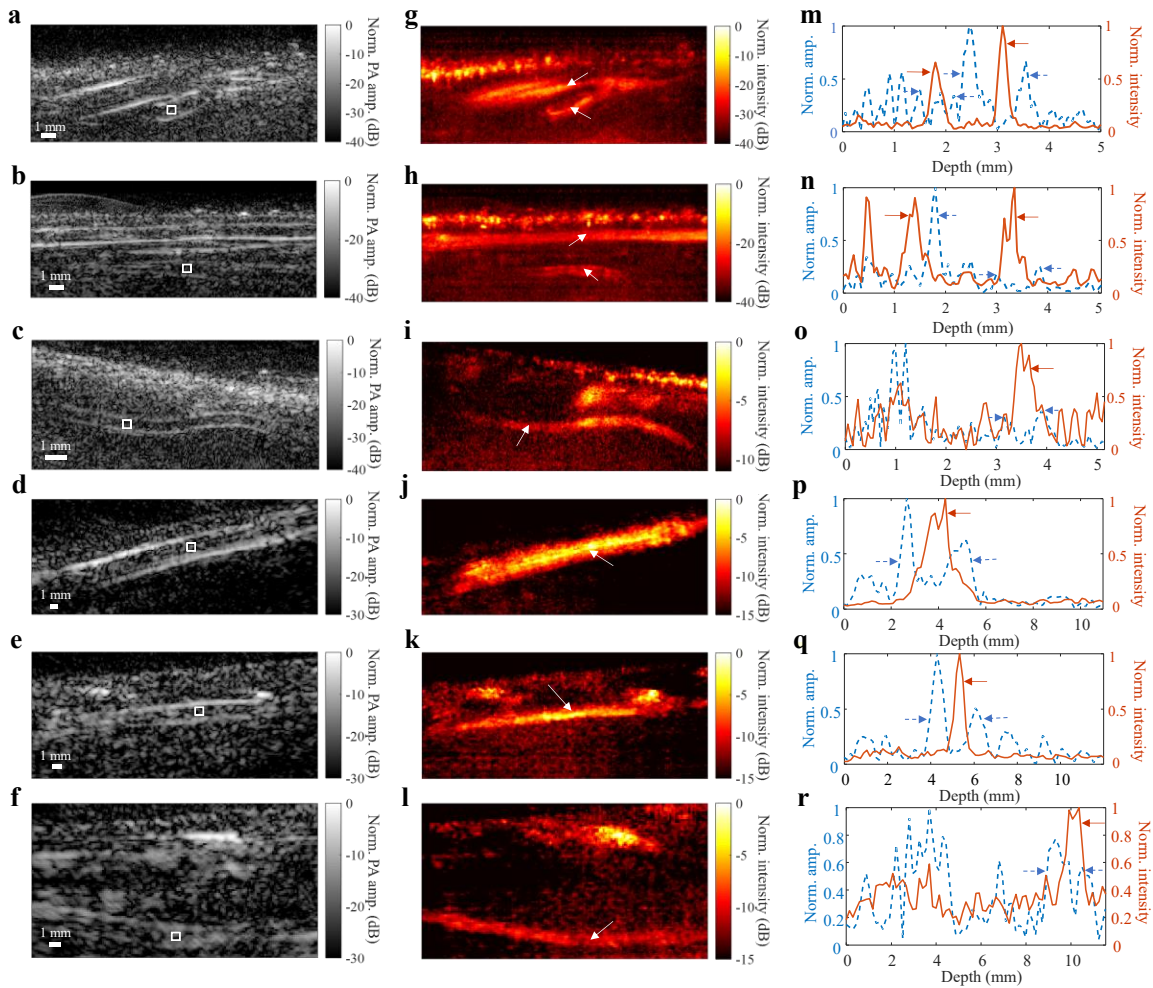

**Fig. 3 | In vivo PANDA images. a-f,** Structural PACT images. **g-l,** Power Doppler PANDA images corresponding to structural counterparts in **a-f**, respectively. Note that the locations of the vessels are as follows: **a-b** represents veins in the carpal tunnel region, **c** represents a digital artery, and **d-f** represent veins in the forearm regions. **m-r,** Vertical profiles of the structural and power Doppler images. Here, depth is measured relative to the skin surface. Location references for the profiles are given by the white boxes in **a-f**, whereby for both the structural and power Doppler images, vertical profiles were drawn according to the central location of the white boxes. The dashed blue lines represent the profiles of the structural images, whereas the solid orange lines represent the profiles of the power Doppler images. The white arrows in **g-l** indicate the vessels of interest in the power Doppler images. The blue and orange arrows in **m-r** indicate the vessel boundaries and vessel lumen regions in the structural PACT and power Doppler images, respectively.

We formed a 3D map of the vasculature in the carpal tunnel region of the arm by scanning the ultrasonic probe along its elevational dimension. All the vessels in the image were located at depths beyond the optical diffusion limit. Three different orientations of the vasculature network are shown in Fig. 4. In the first orientation (Fig. 4a-b), the four vessels indicated by the white arrows in the structural PACT image have hollow lumen regions. Conversely, in the corresponding power Doppler image, these lumen regions are better delineated by filling in the lumen regions. In the second orientation (Fig. 4c-d), there are three vessels indicated by the white arrows whose boundaries have poor contrast relative to the background in the structural image; however, in the corresponding power Doppler image, all three of these vessels are more clearly delineated. The third orientation (Fig. 4e-f) shows that the power Doppler image enhances the visibility of two regions (indicated by the white boxes) of the vasculature that are only weakly detectable in the corresponding structural image. Overall, the power Doppler images show better delineation of the blood vessels compared to the structural images, with cleaner background, less blurring, and better vessel contrast. Most importantly, the power Doppler images provide blood flow information of the vasculature that is absent in the structural PACT images.

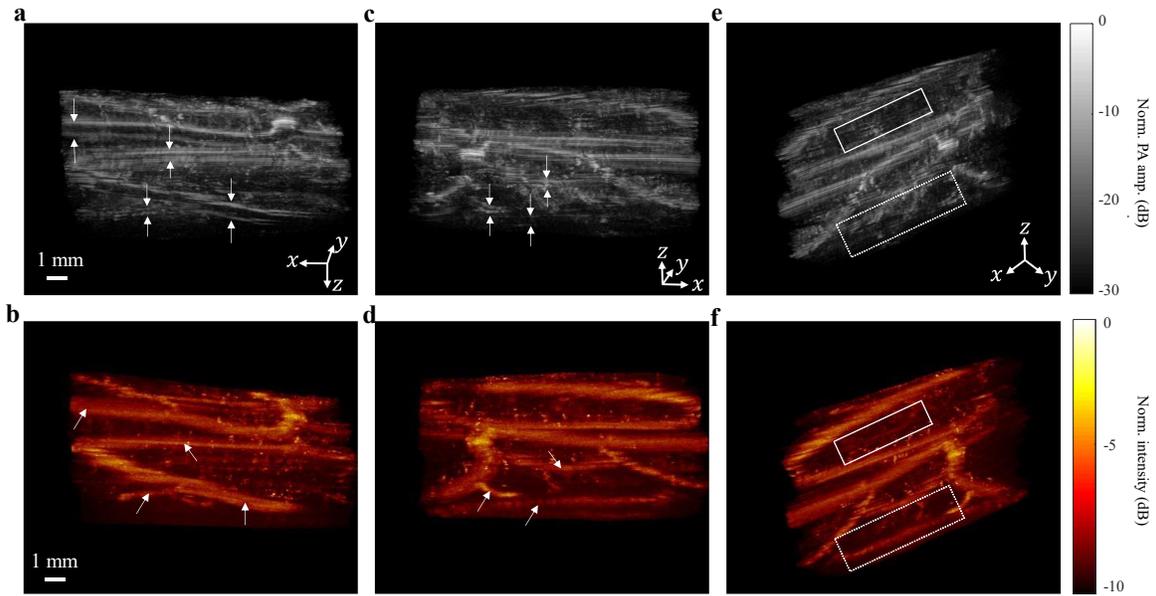

**Fig. 4 | In vivo 3D PANDA images acquired using a linear array. a,c,e,** Structural PACT images of the vasculature in the carpal tunnel region of the arm in orientations 1, 2, and 3, respectively. **b,d,f,** Power Doppler images corresponding to the structural counterparts in **a,c,e** respectively. White arrows indicate the vessel boundaries in **a,c** and vessel lumens in **b,d**. White boxes indicate the two regions for vasculature comparison in structural PACT (**e**) and PANDA (**f**). Note: all coordinates are right-handed.

In addition to scanning a 1D linear array to form a 3D map, we employed a 2D matrix array with a 7.5 MHz center frequency to form fast, successive 3D volumetric functional images of the human vasculature. Representative PANDA projection images ($10\times10\times10$ mm$^3$ volumes) of the carpal tunnel and dorsal hand are shown in Fig. 5a-c and Fig. 5d-f, respectively. For each subfigure, the structural PACT and functional PANDA images are shown in the top and bottom panels, respectively, with the volume displayed at two projected orientations (left vs. right panels, respectively). The left panels (orientation 1) demonstrate that both structural PACT and PANDA can obtain images of the vasculature. The right panels (orientation 2) show that whereas structural PACT images the relative hemoglobin contents at the blood boundaries (i.e., hollow-appearing structures indicated by the white arrows), PANDA can capture the blood flow information from the entire lumen of the vessels.

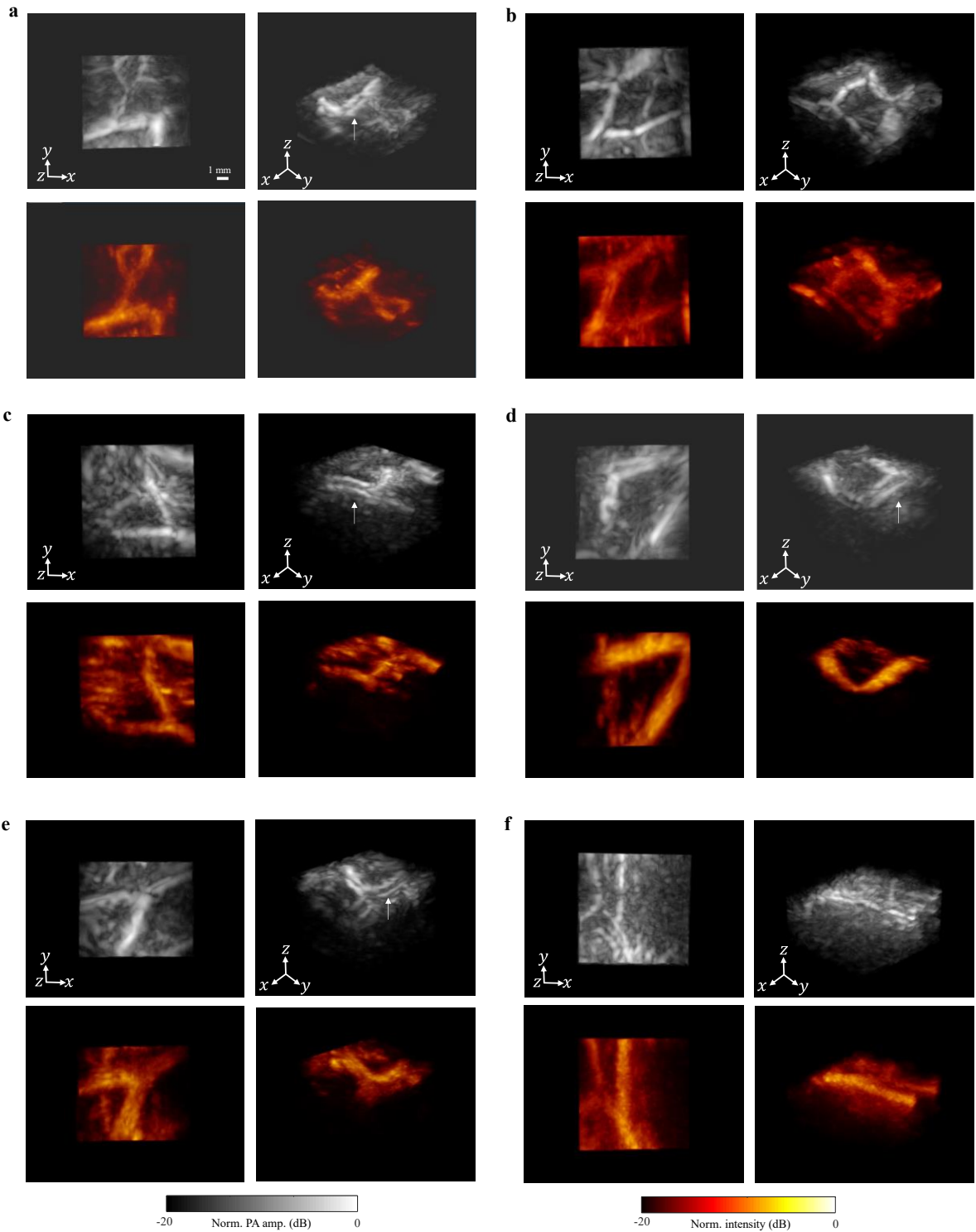

**Fig. 5 | In vivo PANDA images acquired using a matrix array.** Successive 3D images were temporally compiled for six regions in the upper extremities. Structural PACT (top panels) and PANDA (bottom panels) images displayed in orientations 1 (left panels) and 2 (right panels) acquired in the carpal tunnel (**a-c**) and dorsal hand (**d-f**).

To measure PANDA functional changes, we imaged a subject's blood flow before, during, and after inflating a blood pressure cuff. The cuff was applied to the subject's brachial vein, and we imaged a distal vessel in the forearm region (as shown in Fig. 6a). The structural PACT image of the distal vessel is shown in Fig. 6b. Baseline flow was measured for approximately 10 seconds, and fractional changes were calculated relative to this baseline in subsequent measurements. The baseline, cuffed, transient release, and steady-state release power Doppler and color Doppler images are shown in Fig. 6c-j, respectively. In Fig. 6k-l, the time-discretized fractional changes for the power Doppler and color Doppler maps, respectively, both show a flow decrease of nearly 100% while the cuff was inflated, followed by a transient increase of approximately 100% upon release of the cuff and eventually a steady-state return to the baseline. The transient increase could be due to an immediate pressure release in the forearm vessel. A similar trend is confirmed in Supplementary Fig. 6, whereby we performed additional cuff trials on different subjects.

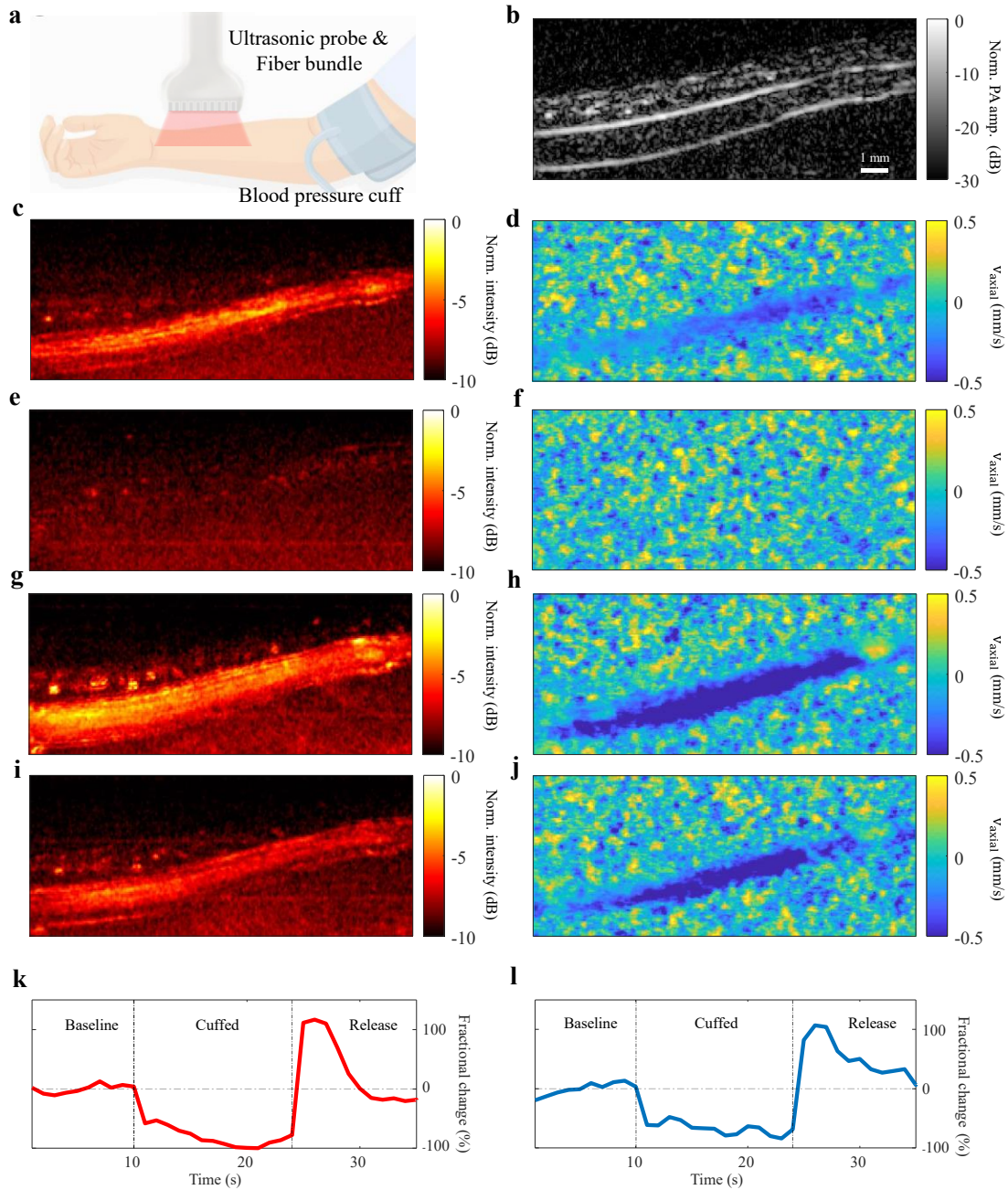

**Fig. 6 | Functional PANDA responses to a blood pressure cuff. a,** Schematic for the experimental setup. A blood pressure cuff was applied to the brachial region of the arm while imaging was performed in the forearm region distal to the cuffing site. **b,** Structural PACT image of a vein in the forearm region. **c,e,g,i,** Power Doppler PANDA images during baseline, cuffing, transient release, and steady state release, respectively. **d,f,h,j,** Color Doppler PANDA images during baseline, cuffing, transient release, and steady state release, respectively. Here, the negative speed is representative of the flow moving away from the probe (i.e., right to left). **k,l,** Fractional changes plotted against time for power Doppler and color Doppler maps, respectively. For both the power Doppler and color Doppler maps, during the cuffing period, the measured values dropped by approximately 100%, after which releasing the cuff caused a transient increase of approximately 100% before settling to steady-state baseline levels.

We formed a 3D map of the vasculature in a patient presenting with varicose veins by imaging the dorsal region of the foot. Varicose veins are characterized by a twisted pattern that can result in swelling, blood vessel damage, and poor circulation[21]. Two different orientations of the vascular network are shown in Fig. 7 with the structural PACT and functional PANDA images in Fig. 7a-b and Fig. 7 c-d, respectively. Zoomed-in views of both the structural PACT (Fig. 7e-f) and PANDA (Fig. 7g-h) images reveal a twisted vascular pattern, with the PANDA images additionally revealing a clear delineation of the irregular functional blood volume distribution. This tortuous pattern is particularly prevalent in the vessels indicated by the white arrows in Fig.7 g-h.

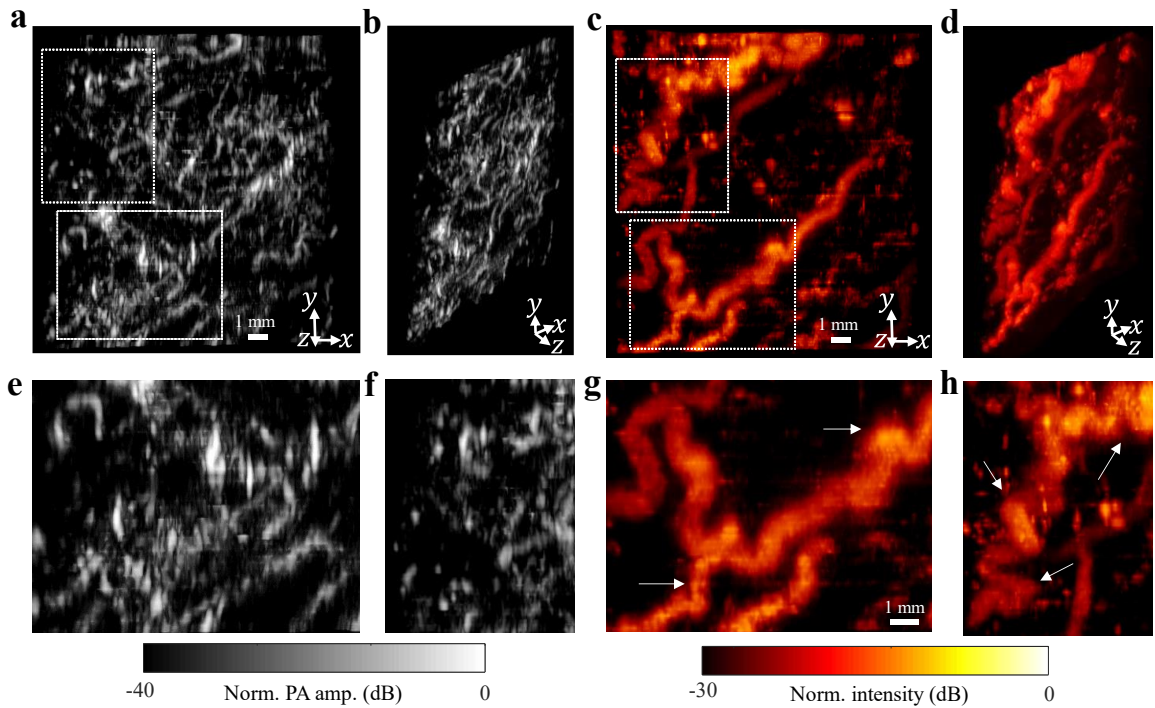

**Fig. 7 | In vivo 3D PANDA images acquired on a patient with varicose veins. a-b**, Structural PACT images of the vasculature at orientations 1 and 2, respectively. **e-f**, Zoomed in views of the regions indicated by the white boxes in **a**. **c-d,g-h,** Power Doppler images corresponding to the structural counterparts in **a-b,e-f**, respectively.

## Discussion

In this article, we introduce PANDA as a physically robust technique for measuring blood flow in humans at depths up to ten times beyond the optical diffusion limit. We confirm the validity of PANDA through *ex vivo* phantom experiments, wherein we demonstrate a proportional relationship between measured photoacoustic Doppler measurements and preset syringe flow speeds. Furthermore, we implement PANDA *in vivo* to characterize both the 2D and 3D blood volume dynamics of the human vasculature. In addition to providing blood flow measurements, we show that PANDA images can offer better vessel delineation than their structural PACT counterpart images, with fuller lumen coverage and stronger contrast being notable features. We also employed a matrix array probe to acquire fast, successive 3D volumetric images in the carpal tunnel and dorsal hand regions, thus demonstrating that PANDA is capable of four-dimensional blood volume functional mapping of the vasculature. Under this implementation, we achieved an effective volumetric imaging rate of 2.5 Hz, whereas the linear-array-based implementation necessitates scanning periods on the order of minutes. Functional imaging of blood flow changes beyond the optical diffusion limit is further demonstrated by measuring the *in vivo* flow responses to a blood pressure cuff, and the imaging of a patient with varicose veins demonstrates the viability of PANDA for immediate clinical impact. Most importantly, PANDA offers a portable and noninvasive solution for providing deep blood flow information that is not present in PACT images, thus making it a potentially powerful clinical method for functional imaging in humans.

For many years, PACT has been utilized for imaging deep-tissue vasculature and has recently shown promise for the diagnosis and monitoring of disease[22,23]. However, due to its lower resolution in comparison to optical resolution photoacoustic microscopy (OR-PAM), PACT is incapable of resolving individual red blood cells (RBCs), whose diameters range from 7-8 microns[24]. Additionally, the random summation of absorption signals from millions of RBCs in each imaging voxel results is considerably weaker for photoacoustic lumen signals relative to the signals at the blood boundaries[25]. As such, measuring blood flow in deep tissue has long been considered beyond the capability of PACT. Based on our point-source phantom experiments, the best resolution that we achieved for our probe was 125 μm and 150 μm in the axial and lateral dimensions, respectively. We note that these values are more than 10 times larger than the diameter of a red blood cell; thus, we are not resolving the propagation of single red blood cells, but rather the composite interference signal from groups of cells. In this work, we hypothesize that the detectability of flow via PANDA is enhanced by the spatial heterogeneity of blood that is resolvable at our probe's center frequency. To test this hypothesis, we designed a flow phantom consisting of a converging bifurcation with two inlet channels and an outlet channel. As shown in Supplementary Fig. 3, we were unable to detect flow when uniform blood was perfused through a single channel. However, by inducing pressure fluctuations in the inlet channels of the bifurcation phantom, we were able to measure blood flow in the lumen region of the outlet channel (see Fig. 2). In the resultant color Doppler and power Doppler maps, the axial velocity and intensity measurements, respectively, were validated against five preset syringe flow speeds, demonstrating a proportional relationship in both cases.

Compared to structural PACT images, PANDA offers the following advantages. 1) PANDA offers additional functional blood flow contrast, which is a critical physiological parameter for assisting in medical diagnoses and treatments; 2) Whereas in structural PACT, only the blood boundaries are highlighted due to the boundary buildup effect[25], PANDA highlights the entire lumen region of the blood vessel (see Fig. 3,4,7); 3) PANDA overcomes the limited view problem[26], a phenomenon in which structural PACT blood boundaries whose surface normal vectors point orthogonally to the detection surface normal cannot be accurately reconstructed (e.g., the side blood boundaries of the structural PACT image in Fig. 2c are missing compared with the fully connected circular lumen regions of the PANDA images in Fig. 2f); 4) By highlighting the dynamic blood flow components, PANDA displays much less tissue background and better vessel delineation than structural PACT (see Fig. 3,4,7).

Compared to other photoacoustic blood flow estimation techniques, such as photoacoustic vector tomography[27] (PAVT), PANDA differs in its mechanism, namely, by utilizing the Doppler effect as a physically robust parameter for flow measurement. Both PAVT and PANDA have unique strengths and weakness. For instance, our phantom results show that power Doppler intensity values are proportional to the preset syringe flow speeds with the vessel positioned in both longitudinal and cross-sectional orientations (which PAVT is incapable of, as shown in Supplementary Fig. 9). This new capability offers a clinically viable solution for measuring relative changes in blood flow with a linear array probe (i.e., in instances where only 2D cross-sectional slices are accessible). It is precisely this technical advance which enabled us to implement a scanning mechanism to obtain 3D angiographic images of blood flow in the human vasculature (Fig. 4 & 7) for which PAVT is currently incapable of. The matrix-array-based implementation of PANDA also potentially enables four-dimensional hemodynamic imaging. Furthermore, in addition to venous flow, PANDA allows us to clearly visualize cuff-induced arterial flow, which has yet to be demonstrated in PAVT. In its current power Doppler implementation, PANDA detects flow at centimeter-level depth, whereas PAVT achieves only millimeter-level detection; this is a consequence of PANDA's temporal integration of multiple frames, whose cumulative effect is potentially more sensitive to weak signals, whereas PAVT relies on tracking frame-to-frame blood motion. For example, as shown in Supplementary Fig. 10, we have obtained the PANDA and PAVT images of a blood vessel in the dorsal region of the hand with overlaying chicken tissue at depths of 1 mm, 3 mm, 5 mm, and 7 mm, respectively. The contrast-to-noise ratios (CNRs) of PANDA (PAVT) are 654 (49), 73 (20), 41 (12), and 20 (4) at depths of 1, 3, 5, and 7 mm, respectively. Compared with PAVT, we also designed our PANDA system to be portable, and we demonstrated its clinical feasibility by forming a 3D blood flow map in a varicose veins patient. Based on the above advantages, PANDA offers the capability to provide more clinically optimal and comprehensive medical diagnoses. However, PANDA in the color Doppler mode maps only the projection of the velocity along the acoustic axis and in the power Doppler mode maps the relative blood flow, whereas PAVT maps the true flow speed and direction. Therefore, PANDA and PAVT are complementary for photoacoustic blood flow imaging.

From the imaging system perspective, we implemented new hardware (e.g., low center frequency ultrasonic detection) to obtain flow maps at depths which exceed the optical diffusion limit by an order of magnitude. Achieving centimeter-level depth opens up the possibility of applying our blood flow techniques to clinically relevant scenarios, such as the jugular vein and carotid artery, breast cancer imaging, and cortical function in the brain. Moreover, we built the PANDA imaging system onto a mobile platform to enable portability for fast clinical translation so that it may be used ubiquitously in the hospital for bedside, intraoperative, and other real-time monitoring.

One important aspect of PANDA is the implementation of spatiotemporal filtering. The conceptual motivation behind the spatiotemporal filtering employed in this work is illustrated in Supplementary Fig. 4. Briefly, we perform singular value decomposition (SVD)[28] on our space-time images, after which we remove the static or slow moving tissue and background noise components corresponding to the lowest and highest singular value indices, respectively, thus allowing us to extract the dynamic blood components from the images. As shown in Supplementary Fig. 5, the user's choice of upper and lower singular value cutoff indices can affect the blood vessel delineation. For example, choosing a lower cutoff value that is too small will leave residual static tissue components in the power Doppler image, whereas choosing an upper cutoff value that is too high will leave background noise in the image, thus making it difficult to properly detect all of the features. To determine the threshold of the singular values, we started by removing the first few large singular values and tested its performance for the removal of the tissue components; we then chose the threshold that successfully removed the tissue components. Further work is needed to implement adaptive algorithms[29] for determining the optimal singular value cutoff indices. Moreover, the filtering process is not limited to an SVD-based implementation. Reasonable Doppler maps may also be obtained from temporal high-pass or bandpass filters.

As a hybrid modality, PANDA is uniquely positioned in the field of biomedical imaging for hemodynamic quantification. Compared to existing pure optical methods, PANDA can obtain Doppler maps in vasculature located at depths beyond the optical diffusion limit. The full power of PANDA, however, comes from its intrinsic optical absorption contrast. Whereas ultrasound can assess blood flow and the elastic properties of biological tissue, PANDA offers the capability to merge endogenous hemoglobin-based optical absorption contrast with flow measurements to quantify blood flow and tissue oxygenation simultaneously, as shown in the Supplementary Fig. 8. We have provided the Doppler ultrasound validation of PANDA blood flow measurement, as shown in Supplementary Fig. 7. The physical meaning of the PANDA structural image (Supplementary Fig. 7a) is dominated by the relative oxy-hemoglobin content of the blood vessel because we used a 1,064 nm laser. The structural ultrasound image (Supplementary Fig. 7d) shows the acoustic backscattering properties of the soft tissue. In the representative color Doppler images, the photoacoustic flow speed map (Supplementary Fig. 7b) agrees with that of the Doppler ultrasound (Fig. 7e). For the power Doppler maps (Supplementary Fig. 7c and f), both imaging modalities show high intensity values in the blood lumen regions, which indicates a higher blood volume than that of the tissue region. The congruence between the two modalities is further conveyed in representative vertical profiles from the structural, color Doppler, and power Doppler (Supplementary Fig. 7 d-f, j-l).

The spectral implementation of PANDA broadens its capabilities to simultaneously measure blood flow, total hemoglobin concentration (HbT), and oxygen saturation (sO$_2$). Together, these parameters can provide a more comprehensive physiological profile for clinicians. As shown in Supplementary Fig. 8, PANDA provided simultaneous measurement of hemoglobin oxygen saturation ($sO_2$) and the blood flow. Furthermore, future implementations of PANDA may extend its scope to metabolic imaging of physiological systems. The simultaneous measurements of blood flow, HbT, and $sO_2$ in the major arteries and veins of a given system can be used to quantify the metabolic rate of oxygen consumption ($MRO_2$)[30]. As shown in Figure 3i, we were able detect flow in a digital artery by applying a proximal cuff to the brachial arm region. However, in order to reach the full clinical potential of PANDA in metabolic imaging, future work will need to explore a more robust implementation of the technique whereby baseline arterial flow is detectable.

We employed PANDA using a linear array probe to image blood flow in the peripheral vasculature. Future work may extend the scope of this technique by using other array geometries for broader clinical applications. For instance, full-view detection from a ring array would provide enhanced imaging quality[31], while a hemispherical array may provide a larger field-of-view for 3D volumetric imaging of the vasculature[32]. Such enhancements could facilitate the application of PANDA to brain functional imaging[7,33], early breast cancer detection of tumor-mediated hypermetabolism[34], and metabolic imaging of whole organ systems.

In this work, we first extracted Doppler information for blood flow beyond 1 millimeter, from many vessels ranging from 1 to 10 millimeters in depth, thus surpassing a longstanding impasse in the field of biomedical optics. This paves the way for imaging even deeper: since diffuse photons can generate photoacoustic signals at depths up to several centimeters, we may expect to achieve deeper measurements by employing lower frequency probes in future studies. Here, we primarily imaged the upper and lower extremities, regions which may already have clinical significance to individuals with peripheral vascular diseases. In particular, PANDA may assist physicians in the hemodynamic characterization of oxygenation and blood flow in individuals suffering from varicose veins, an affliction which can manifest in patients who are overweight, pregnant, or diabetic. Early diagnosis of varicose veins enabled by PANDA may help physicians to develop proactive treatment plans to prevent further progression of circulatory complications, such as skin ulcers, necrosis, and blood clots[21]. We chose the upper and lower extremities regions as proof-of-concept demonstrations, but our methods can certainly be applied to other clinically relevant regions of the circulatory system to diagnose vascular function and pathology at centimeter-level depths. Examples include diabetes-induced pedal vascular trauma, the jugular vein and carotid artery, intraoperative monitoring of incision-adjacent tissue perfusion, and hypermetabolism in breast tumors.

PANDA synergizes the photoacoustic and Doppler effect to provide centimeter-level imaging of both hemoglobin and blood flow. Whereas traditional photoacoustic computed tomography and Doppler ultrasound are limited to exclusively measuring hemoglobin content or blood flow, respectively, PANDA is uniquely positioned to provide dual-contrast, noninvasive, three/four-dimensional images using a portable, single modality for patients. Taken as a whole, these distinctive features of PANDA potentially enable comprehensive functional diagnoses that are thus far unparalleled in medicine.

In its current state, PANDA is primed for rapid clinical translation. By extending the achievable depth for measuring blood flow to one order of magnitude higher than the optical diffusion limit, PANDA reveals the deep hemodynamics of human vasculature previously considered inaccessible to optical imaging. We also designed PANDA to be integrated into a portable system, and this mobility enables our system to be transported to the hospital for bedside implementation. Thus, PANDA may emerge as a powerful functional imaging modality for aiding clinicians in diagnosing and preventing disease in patients.

**Methods**

**System construction.**

For the detection of photoacoustic signals, we utilized a 256-element linear ultrasonic transducer array (LZ250, VisualSonics Inc.) with a center frequency of 15 MHz, a 1,024-element matrix array (32 × 32, 0.3 mm pitch, Vermon, Verasonics, Inc.) with a center frequency of 7.5 MHz, and a 128-element linear array (ATL L7-4, Philips) with a center frequency of 5 MHz. Unless otherwise specified (as in Fig. 3d-f and Fig. 5), all images in this manuscript were acquired with the LZ250 probe. The ultrasound probe was linked via a UTA 360, UTA 1024-MUX, or UTA 260-D connector to the Verasonics Vantage 256 system (Verasonics Inc.), which has 256 DAQ channels, 14-bit A/D converters, a 60 MHz, 30 MHz, or 20 MHz sampling rate for the LZ250, matrix probe, and ATL L7-4 probes, respectively, and programmable gain of up to 51 dB. The photoacoustic signals were captured and converted into digital form, which were then saved in local memory and transferred to a host computer via PCI express.

We used a 1,064 nm wavelength laser beam to pass light through an optical fiber bundle to the imaging target. The fiber bundle was coupled to and coaxially aligned with the ultrasound probe. The lasers were operated at either a 10 Hz pulse repetition frequency (PRF) (Quantel Brilliant B pulsed YAG laser; 5-6 ns pulse width), 20 Hz PRF (Quantel Q-Smart 450 laser; 5-6 ns pulse width), or 100 Hz PRF (SpitLight EVO III, InnoLas Laser Inc.; 5-8 ns pulse width). The optical fluences were approximately 50 mJ/cm$^2$ for the 1,064 nm wavelength at 10 Hz PRF, 30 mJ/cm$^2$ at 20 Hz PRF, and 6 mJ/cm$^2$ at 100 Hz PRF, which were less than the safety limit set by the American National Standards Institute (ANSI) (100 mJ/cm$^2$, 50 mJ/cm$^2$, and 10 mJ/cm$^2$, respectively, and 1000 mW/cm$^2$)[35].

We triggered the DAQ for signal acquisition using the laser's external trigger. To correct for the delay and jitter between the DAQ and the laser, we acquired the signal from the ultrasound probe surface and temporally aligned the maximum of this signal with the first sample of the DAQ acquisition sequence in each imaging frame. The preprocessed raw signals were then quadrature sampled and back-projected to reconstruct the two-dimensional (2D) photoacoustic image.

**Data acquisition.** For human imaging, the subject's arm was placed inside a portable tank below the ultrasound probe; the imaging target and the probe surface were acoustically coupled by filling the tank with water. We used an imaging frame rate of 10 Hz, 20 Hz, or 100 Hz corresponding to the laser PRF. To obtain 3D flow images, we mounted the probe to a motorized translation stage (PI Micos) and linearly scanned the probe along its elevational dimension. The motor was synchronized to the laser's external trigger. For Fig. 5, the motor was operated at a step size of 100 microns. The total scan acquisition time was approximately 120 seconds. For Fig. 5., the images were acquired with a 10 Hz laser and multiplexed 4:1 for an effective volume rate of 2.5 Hz (total reconstruction volume dimensions 10 mm×10 mm×10 mm). For Fig. 7, the motor was operated at a step size of 250 microns with 50 frames acquired at each step at a laser pulse repetition frequency of 20 Hz, and the total scan range and acquisition time were approximately 25 mm and 250 seconds, respectively. In the blood pressure cuff functional experiment, a blood pressure cuff device (GF Health Products, Inc) was placed on the upper arm region. The total acquisition time was approximately 35 s, consisting of a baseline measurement of ~10 s, followed by a cuff period of ~15 s, and a release period of ~10 s.

A flow phantom was assembled from micro-renathane tubing (BrainTree Scientific). Each inlet channel had an inner diameter of 0.6 mm, and the outlet channel had an inner diameter of 1.0 mm. The phantom was perfused with 45% hematocrit whole bovine blood (QuadFive). Pressure fluctuations were induced by gently squeezing the inlet channels.

Ultrafast ultrasound Doppler imaging was implemented by coherent plane wave compounding[36,37] with seven tilted plane waves from $-14$ degrees to $+14$ degrees at a pulse repetition frequency of 10,000 Hz. The Doppler information was extracted after applying singular value decomposition (SVD)[28]. In the ultrasound validation experiment, the ultrafast ultrasound data were immediately acquired after the PANDA imaging sequence to ensure the same state in the region of interest.

**Data processing.** To obtain Doppler maps from the acquired photoacoustic signals, we implemented the following procedure for data processing. We first convert the raw photoacoustic signals to their quadratures using the Hilbert transformation and employ a universal back-projection algorithm to reconstruct the structural images. A spatiotemporal filter is then used to extract the blood component from the structural images. The spatiotemporal structure dataset has three dimensions with two spatial axes (ultrasound probe azimuthal direction $x$ and axial direction $z$) and one time axis (slow time $t$). We first reshape the 3D dataset into a 2D space-time matrix $S_{\text{structure}}(x, z, t)$. Then we use SVD[28] to decompose the data matrix as follows

$$S_{\text{structure}}(x, z, t) = \sum_{i=1}^{r} \lambda_i u_i(x, z) v_i^T(t), \qquad (1)$$

where $r$ is the rank of the data matrix, $\lambda_i$ is the $i^{th}$ singular value, $T$ is the conjugate transpose, $u_i(x, z)$ corresponds to the spatial domain, and $v_i(t)$ corresponds to the temporal domain. The static or slow-moving components (i.e., tissue) correspond to the first few larger singular values (i.e., the smallest singular value indices), whereas the noise components correspond to the last few smaller singular values (i.e., the largest singular value indices). Therefore, the relatively fast-moving blood components can be extracted as

$$S_{\text{blood}}(x, z, t) = S_{\text{structure}}(x, z, t) - \sum_{i=1}^{\lambda_l} \lambda_i u_i(x, z) v_i^T(t) - \sum_{i=\lambda_u}^{r} \lambda_i u_i(x, z) v_i^T(t)$$

$$= \sum_{i=\lambda_l}^{\lambda_u} \lambda_i u_i(x, z) v_i^T(t), \qquad (2)$$

where $\lambda_l$ and $\lambda_u$ are the lower and upper singular values cutoffs, respectively, for extracting the blood component. Finally, the filtered 2D space-time matrix $S_{\text{blood}}(x, z, t)$ is reshaped back to 3D with the same size as the original 3D dataset. This framework is illustrated in Supplementary Fig. 4.

Each pixel in the image contains an in-phase ($I$) and quadrature ($Q$) component and can be mathematically expressed in complex form as

$$S_{\text{blood}}(x, z, t_n) = I_n + jQ_n = A_n e^{j\phi_n}, \qquad (3)$$

where $S_{\text{blood}}(x, z, t_n)$ represents the blood component pixel value with coordinates $(x, z)$ in the $n$th image frame, $A_n = \sqrt{I_n{}^2 + Q_n{}^2}$ is the envelope of the signal, and $\phi_n = 2\pi f t_n$ is the phase of the received signal with frequency $f$ at time $t_n$. The mean Doppler frequency shift $\bar{f}$ can then be estimated for each pixel from the mean phase shift $\overline{\Delta\emptyset}$ according to[38]

$$\bar{f}(x, z) = \frac{\overline{\Delta\emptyset}(x,z)}{2\pi T_{\text{PRF}}} = \frac{1}{2\pi T_{\text{PRF}}(N-1)} \sum_{n=1}^{N-1} \text{angle}[S_{\text{blood}}(x, z, t_{n+1}) S_{\text{blood}}^*(x, z, t_n)], \quad (4)$$

where $N$ is the total number of frames, $T_{\text{PRF}}$ is the time period associated with the pulse repetition frequency (PRF) of the laser, $^*$ denotes the complex conjugate, and the angle function computes the phase of a complex number $z = x + jy$ as $\text{angle}[z] = \tan^{-1}[\frac{y}{x}]$. Finally, the axial velocity component of each pixel can be obtained from

$$v_{\text{axial}}(x, z) = \frac{c\bar{f}(x,z)}{f_0}, \qquad (5)$$

where $c$ is the speed of sound in the medium, and $f_0$ is the center frequency of the ultrasonic probe. The axial velocity values are then used to construct the color Doppler map. Thus, the color Doppler value is proportional to the blood flow speed.

Alternatively, power Doppler angiography maps can be obtained by calculating the mean intensity of the blood component images according to[38]

$$I_{\text{PD}}(x, z) = \frac{1}{N} \sum_{n=1}^{N} |S_{\text{blood}}(x, z, t_n)|^2, \qquad (6)$$

where $I_{\text{PD}}$ is the power Doppler value, which is proportional to the blood volume. Here, we integrated our signal in the time domain; however, an equivalent integral can be reached in the frequency domain according to Parseval's theorem[39].

Here, we note that for each image in this manuscript, a sole horizontal scale bar was shown in lieu of both a horizontal scale bar and a vertical scale, as the two scale bars are equal in length due to isotropic pixel sampling. For the three-dimensional maps formed by the linear array (i.e., Fig. 4 and Fig. 6), we partially masked out superficial skin signals prior to the final maximum amplitude/intensity projections of the blood vessels. For Supplementary Fig. 10, we acquired 50 frames of interleaved ultrasound and photoacoustic images using a 1,064 nm laser wavelength at a frame rate of 10 Hz. The PANDA images were obtained with a cutoff SVD value set to 15. The processing methods for PAVT in Supplementary Figs. 9 and 10 can be found in reference 26.

**Imaging protocols.**  The experiments on human extremities were performed in a dedicated imaging room. All experiments were performed according to the relevant guidelines and regulations approved by the Institutional Review Board of the California Institute of Technology (Caltech). Ten subjects were recruited from Caltech. Written informed consent was obtained from all the participants according to the study protocols.

**Reporting Summary.**  Further information on research design is available in the Nature Research Reporting Summary linked to this article.

**Data availability**

The data that support the findings of this study are provided within the paper and its supplementary material.

**Code availability**

The reconstruction algorithm and data processing methods can be found in Methods. The reconstruction code is not publicly available because it is proprietary and is used in licensed technologies.

**Acknowledgments**

This work was sponsored by the United States National Institutes of Health (NIH) grants U01 EB029823 (BRAIN Initiative) and R35 CA220436 (Outstanding Investigator Award).

**Contributions**

L.V.W., Y.Z., and J.O.G. designed the study. Y.Z. and J.O.G. built the system, performed the experiments, analyzed the data, and wrote the manuscript with input from all of the authors. K. S. helped in optimizing the laser beam path and configuring the low-frequency system. L.V.W. supervised the study and revised the manuscript.

**Competing interests**

L.V.W. has a financial interest in Microphotoacoustics Inc., CalPACT LLC, and Union Photoacoustic Technologies Ltd., which, however, did not support this work.

**Supplementary figures**

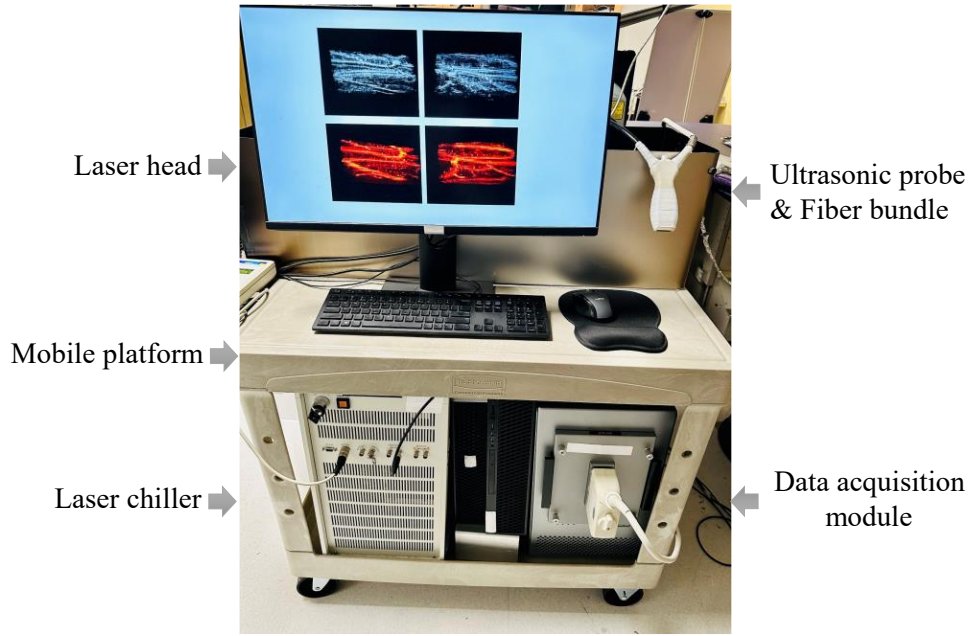

**Supplementary Fig. 1 | Photoacoustic noninvasive Doppler angiography system.** Integration of the laser, optical path, ultrasonic probe, data acquisition module, and host computer onto a mobile platform. The dimensions of the whole system are 110 cm (Length) x 80 cm (Width) x 130 cm (Height).

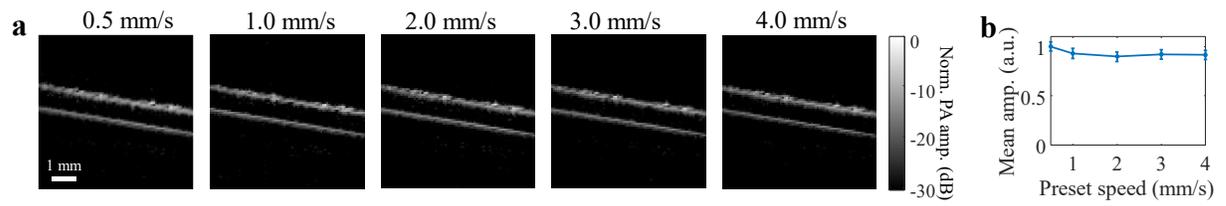

**Supplementary Fig. 2 | PACT image amplitude does not vary with speed. a,** Structural PACT phantom images of the outlet channel across five different speed. **b,** Mean amplitude of the channel wall measured against preset syringe flow speeds. The mean photoacoustic amplitude does not vary with speed.

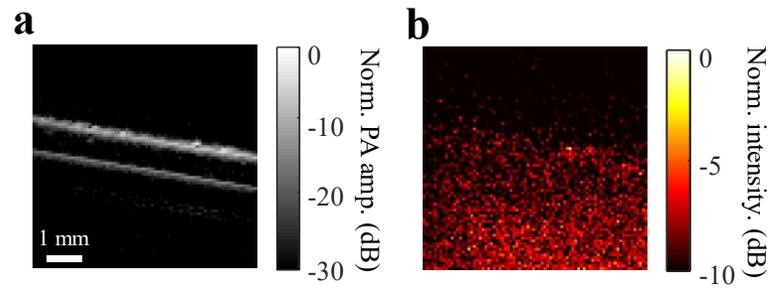

**Supplementary Fig. 3 | Phantom PANDA measurement in a single channel. a,** Structural PACT phantom image of the channel perfused with uniform blood. **b,** Power Doppler map of the channel demonstrates that the flow is undetectable.

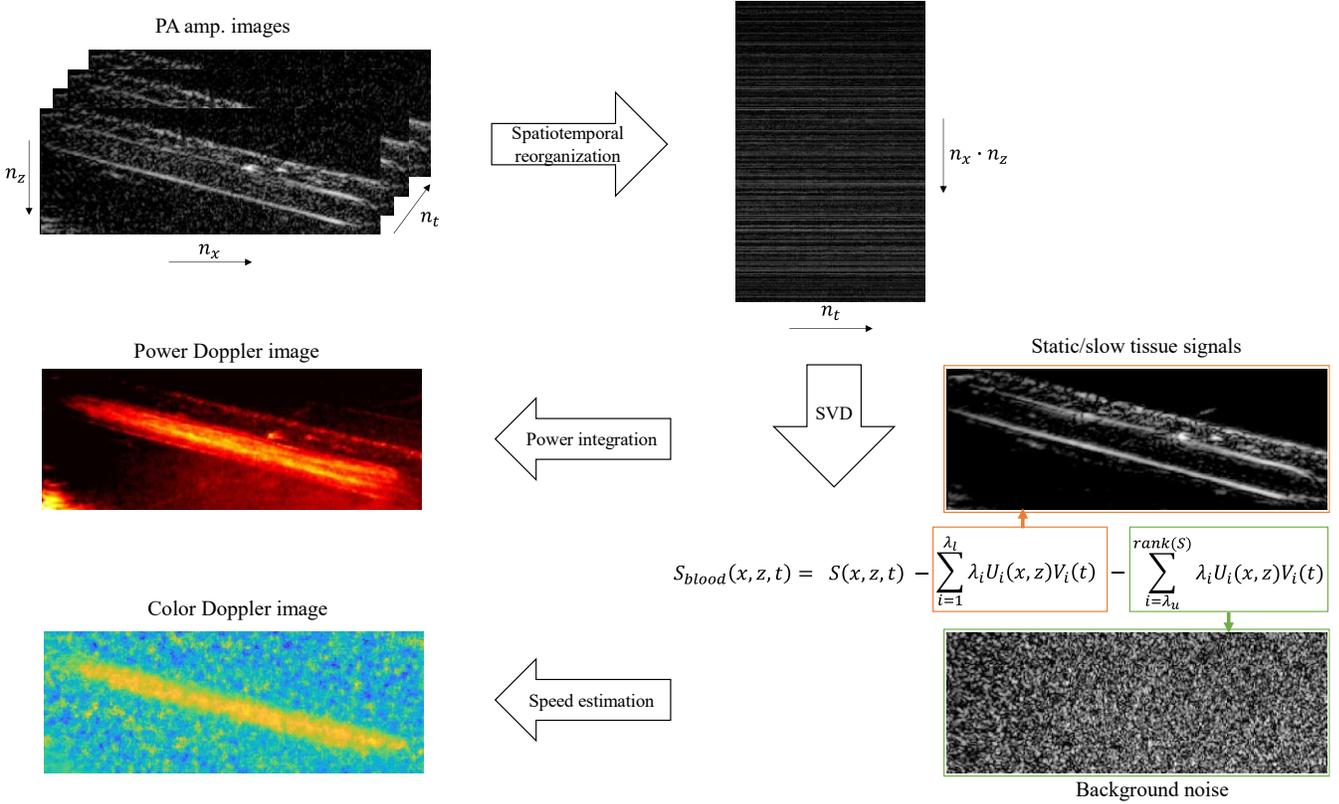

**Supplementary Fig. 4 | Spatiotemporal filtering schematic.** A 2D image stack ($n_z$ x $n_x$) of structural photoacoustic *in vivo* images are acquired across time, forming a 3D spatiotemporal matrix with dimensions $n_z$ x $n_x$ x $n_t$ (top left). The 3D matrix is reshaped to 2D with dimensions $n_x n_z$ x $n_t$ (top right). Singular value decomposition is performed on the reshaped matrix, after which the blood signal is extracted by removing the slow-moving tissue and background noise corresponding to the small and large singular value indices, respectively (bottom right). From the blood signal, power Doppler, and color Doppler images are generated

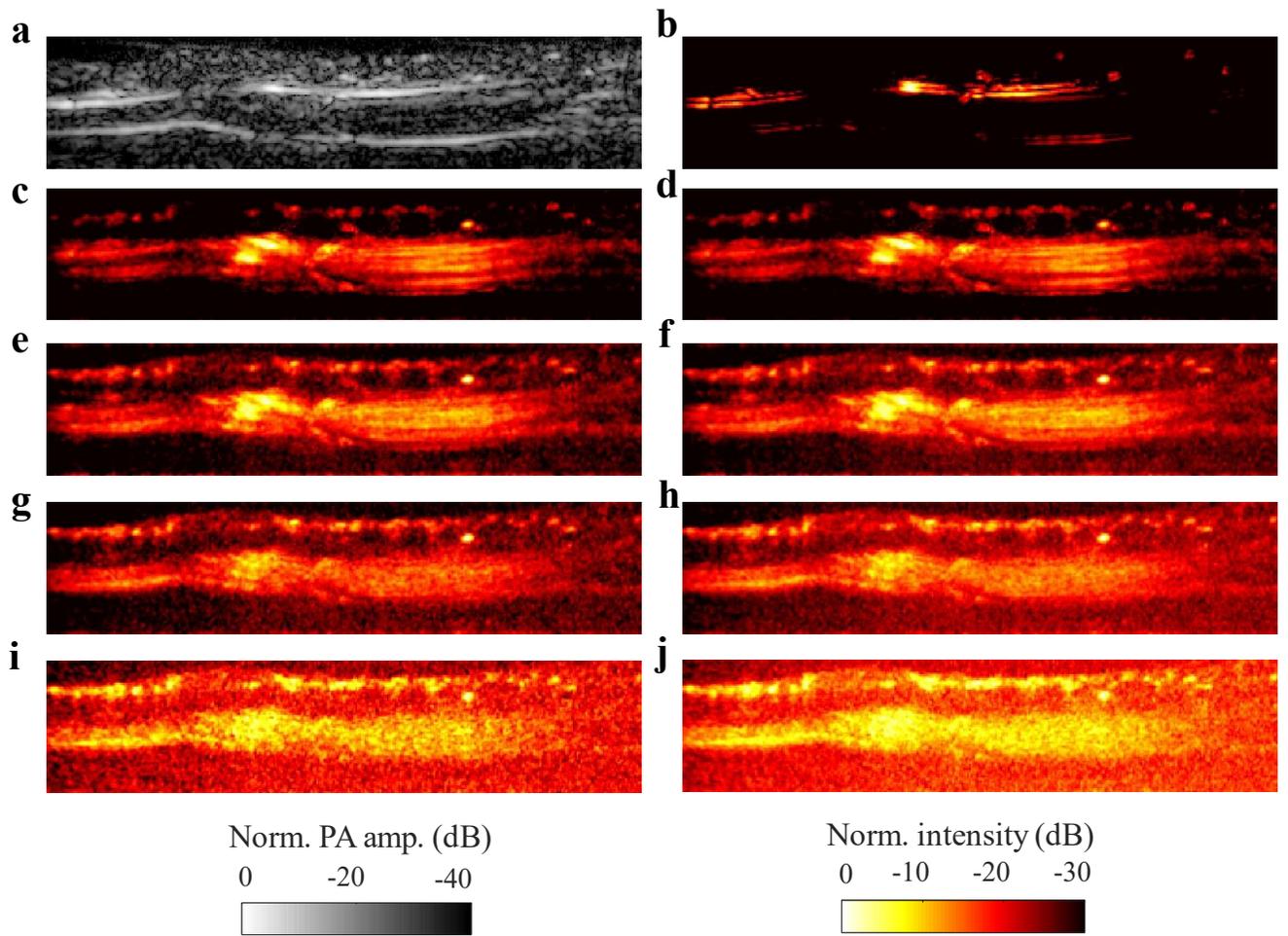

**Supplementary Fig. 5 | Varying the singular value cutoff in PANDA. a,** Structural photoacoustic image of a representative *in vivo* vessel. **b-j,** Resultant power Doppler images were generated from 100 frames of data. **b,d,f,h,j,** Power Doppler images with lower singular value cutoff indices of 1, 10, 20, 40, and 60, respectively. **c,e,g,i,** Power Doppler images with upper singular value indices of 80-100 removed and lower singular value cutoff indices of 10, 20, 40, and 60, respectively.

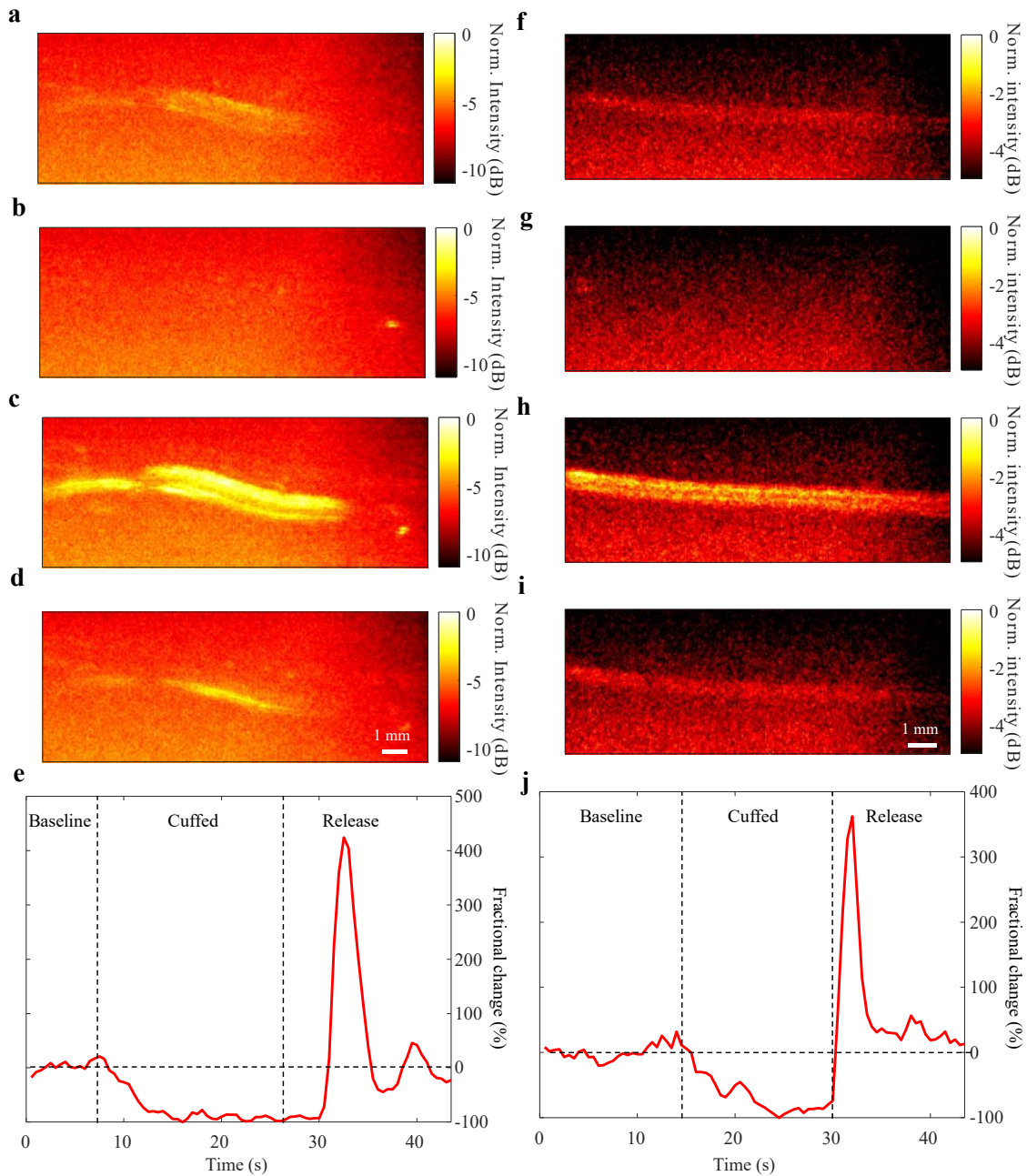

**Supplementary Fig. 6 | PANDA functional response to a blood pressure cuff.** Fractional intensity responses to a blood pressure cuff for different subjects. **a-d**, Power Doppler PANDA *in vivo* images during baseline, cuffing, transient release, and steady state release, respectively, for second subject. **e**, Fractional changes plotted against time for second subject. **f-i**, Power Doppler PANDA *in vivo* images during baseline, cuffing, transient release, and steady state release, respectively, for third subject. **j**, Fractional changes plotted against time for third subject.

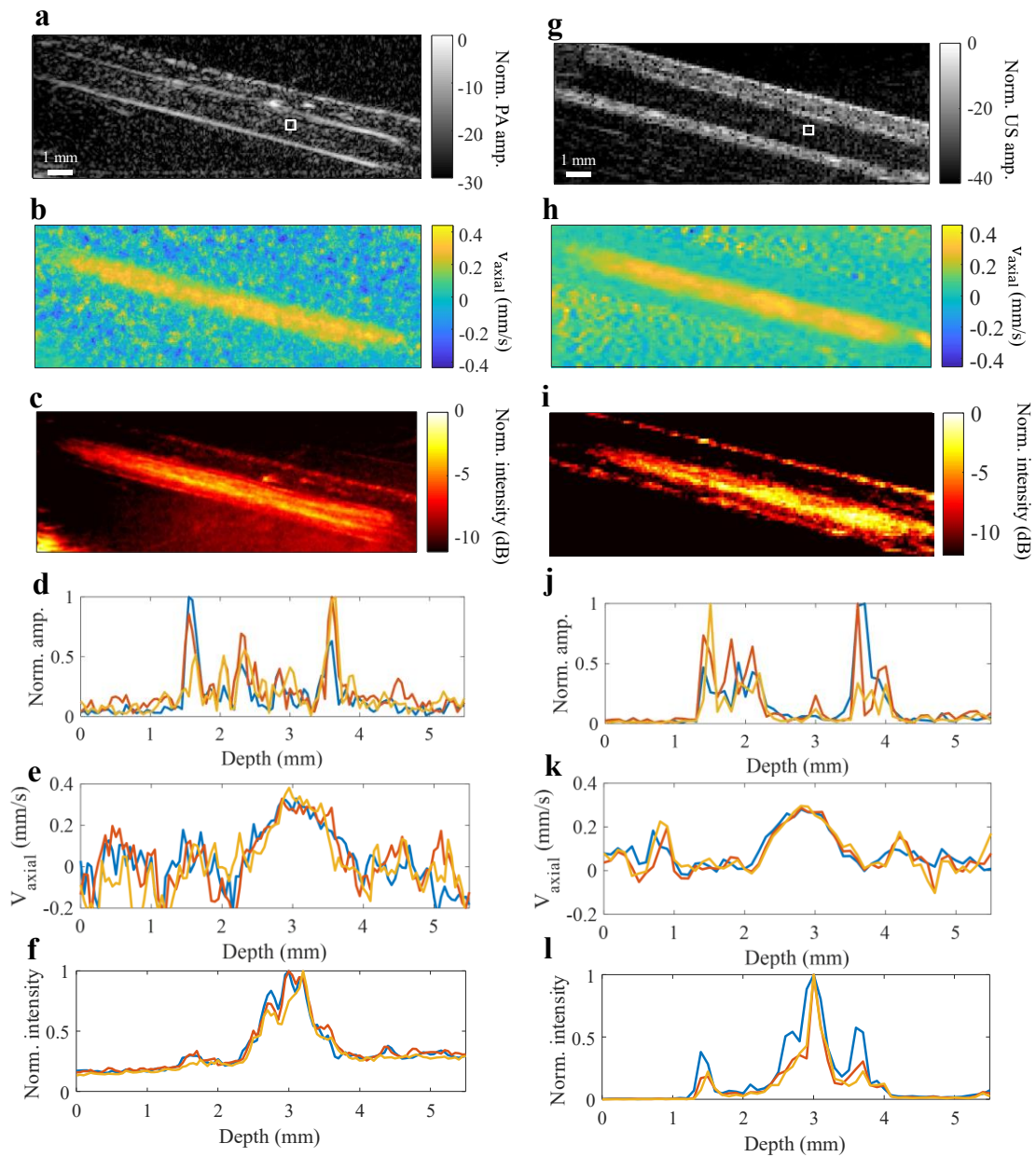

**Supplementary Fig. 7 | In vivo validation of PANDA with Doppler ultrasound. a-c, g-i,** Structural, color Doppler, and power Doppler photoacoustic and ultrasound images, respectively, of a blood vessel in the forearm region. **d-f, j-l,** Vertical profiles corresponding to images in **a-c, g-i,** respectively. Location references for the profiles are given by the white box in **a,** whereby three vertical profiles were drawn according to the locations of three adjacent horizontal pixels in each box, respectively.

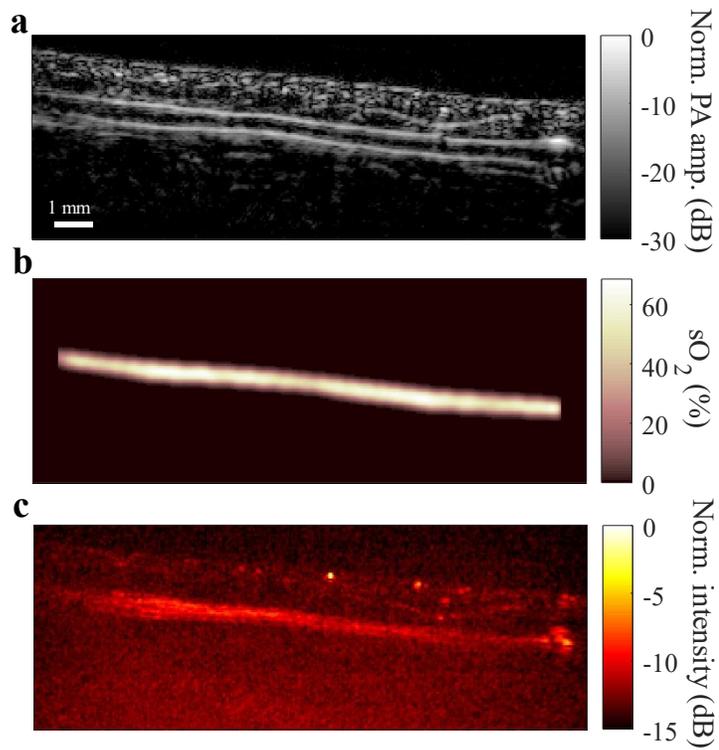

**Supplementary Fig. 8 | Simultaneous measurement of hemoglobin oxygen saturation (sO₂) and relative blood flow in vivo. a,** Photoacoustic structure image (top panel), **b,** sO₂ map (middle panel), and **c,** power Doppler flow map (bottom panel) of three vessels located in the palmar region. The sO₂ map was generated by linearly unmixing images acquired at optical wavelengths of 694, 750, 805, 900, and 1,064 nm. To reject impertinent extravascular signals, masking was performed on the sO₂ map.

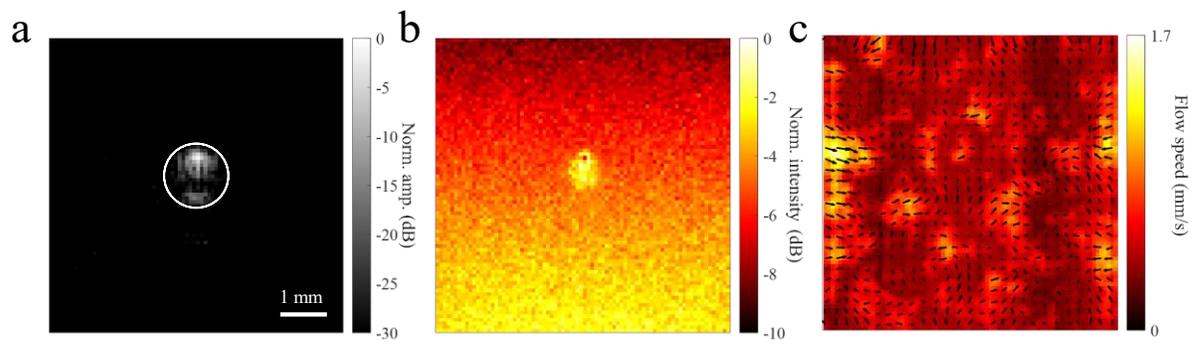

**Supplementary Fig. 9 | Comparison of PANDA with PAVT in a cross-sectional view of a blood tube phantom. a,** Structural photoacoustic image, **b,** PANDA image, and **c,** PAVT image. The ground truth blood flow speed is 4 mm/s. The CNRs of PANDA and PAVT are 9.0 and 0.1, respectively. White circle indicates the region for CNR calculation. PANDA: Photoacoustic noninvasive Doppler angiography; PAVT: Photoacoustic vector tomography.

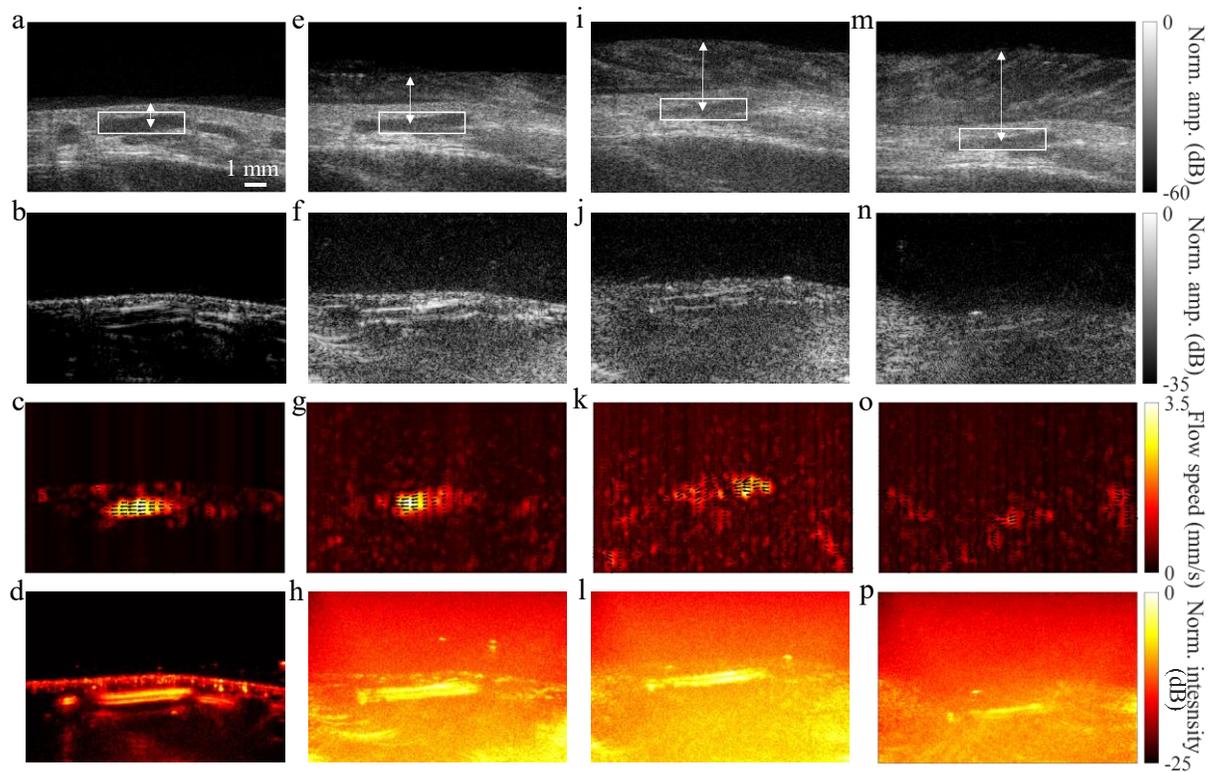

**Supplementary Fig. 10 | Comparison of PANDA with PAVT in vivo at varying depths.** Structural ultrasound image, structural photoacoustic image, PAVT image, and PANDA image of a blood vessel in the dorsal region of the hand with overlaying chicken tissue at depths of 1 mm (**a-d**), 3 mm (**e-h**), 5 mm (**i-l**), and 7 mm (**m-p**), respectively. The CNRs of PANDA (PAVT) are 654 (49), 73 (20), 41 (12), and 20 (4) at depths of 1, 3, 5, and 7 mm, respectively. White arrows indicate the depth of the blood vessel, and the white boxes indicates the regions of interest for the CNR calculation. PANDA: Photoacoustic noninvasive Doppler angiography; PAVT: Photoacoustic vector tomography.